\documentclass[aps,pra,reprint,groupedaddress]{revtex4-1}
\usepackage{graphicx,siunitx}
\usepackage[english]{babel}
\usepackage{braket}
\usepackage{float}
\usepackage{amsmath}
\usepackage{xcolor,soul}
\DeclareMathOperator{\sign}{sign}
\newcommand\qubitup{\ket{\uparrow}}
\newcommand\qubitdown{\ket{\downarrow}}
\newcommand\Ca{$^{40}\text{Ca}^+$}
\newcommand\Sr{$^{88}\text{Sr}^+$}
\newcommand\Ba{$^{138}\text{Ba}^+$}
\newcommand\Ra{$^{226}\text{Ra}^+$}

\begin{document}


\title{A Wavelength-Insensitive, Multispecies Entangling Gate for Group-2 Atomic Ions}


\author{Brian C. Sawyer}
\email[]{brian.sawyer@gtri.gatech.edu}
\author{Kenton R. Brown}
\email[]{kenton.brown@gtri.gatech.edu}

\affiliation{Georgia Tech Research Institute, Atlanta, Georgia 30332, USA}


\begin{abstract}
We propose an optical scheme for generating entanglement between co-trapped identical or dissimilar alkaline earth atomic ions (\Ca, \Sr, \Ba, \Ra) which exhibits fundamental error rates below $10^{-4}$ and can be implemented with a broad range of laser wavelengths spanning from ultraviolet to infrared. We also discuss straightforward extensions of this technique to include the two lightest Group-2 ions ($\text{Be}^+$, $\text{Mg}^+$) for multispecies entanglement. The key elements of this wavelength-insensitive geometric phase gate are the use of a ground ($S_{1/2}$) and a metastable ($D_{5/2}$) electronic state as the qubit levels within a $\sigma^z \sigma^z$ light-shift entangling gate. We present a detailed analysis of the principles and fundamental error sources for this gate scheme which includes photon scattering and spontaneous emission decoherence, calculating two-qubit-gate error rates and durations at fixed laser beam intensity over a large portion of the optical spectrum (300 nm to 2 $\mu \text{m}$) for an assortment of ion pairs. We contrast the advantages and disadvantages of this technique against previous trapped-ion entangling gates and discuss its applications to quantum information processing and simulation with like and multispecies ion crystals. 
\end{abstract}

\maketitle

\section{Introduction}
A variety of entangling gate schemes has been demonstrated in trapped-ion systems. To date, two-qubit gates have been implemented most commonly with two distinct techniques, the light-shift (LS) gate 
\cite{leibfried_experimental_2003,aolita_high-fidelity_2007,ballance_high-fidelity_2016,baldwin_high_2020} and the “Mølmer-Sørensen” (MS) gate \cite{sorensen_entanglement_2000,gaebler_high-fidelity_2016}. While MS gates have been used with optical, hyperfine, and Zeeman transition qubits, LS gates have been implemented only on the latter two. Alternatively, it should be possible perform a light-shift gate with a dipole force on an optical-transition qubit (an ``optical-transition dipole-force" (OTDF) gate), which is less sensitive to several important sources of error encountered in the more common LS or MS gates. A related idea has been explored in Refs.~\cite{roos_ion_2008,kim_geometric_2008} which present a $\sigma^z \sigma^z$ interaction induced via beams at a specific, small detuning near the electric-dipole-forbidden $S-D$ optical transition. 

In contrast, the OTDF gate presented in this Article relies on a state-dependent optical-dipole force and functions in the regime of laser wavelengths detuned far from the $S-D$ transition. Our OTDF gate permits an unprecedented range of feasible laser wavelengths, such that reliable commercial laser systems, including at the telecommunications-relevant 1550 nm wavelength, could be employed. This broad range of wavelengths also makes the OTDF gate straightforward to implement in systems of disparate-species ions~\cite{tan_multi-element_2015,bruzewicz_dual-species_2019,hughes_benchmarking_2020}. In particular, a second species can be used for sympathetic cooling, for measurement of a nearby ion's state without photon-scattering decoherence, or for photonic-qubit wavelength conversion. Another powerful consequence of broad wavelength tuneability is the possibility to tailor the OTDF Lamb-Dicke parameter for a given ion species and trap configuration. As an example, one can choose to minimize the effect of terms outside the Lamb-Dicke approximation when trap confinement is weak or motional cooling is imperfect, or more extreme (i.e. counter-propagating) laser beam geometries can be used in ion traps with constraints on optical access.

\begin{figure}
	\includegraphics[scale=0.65]{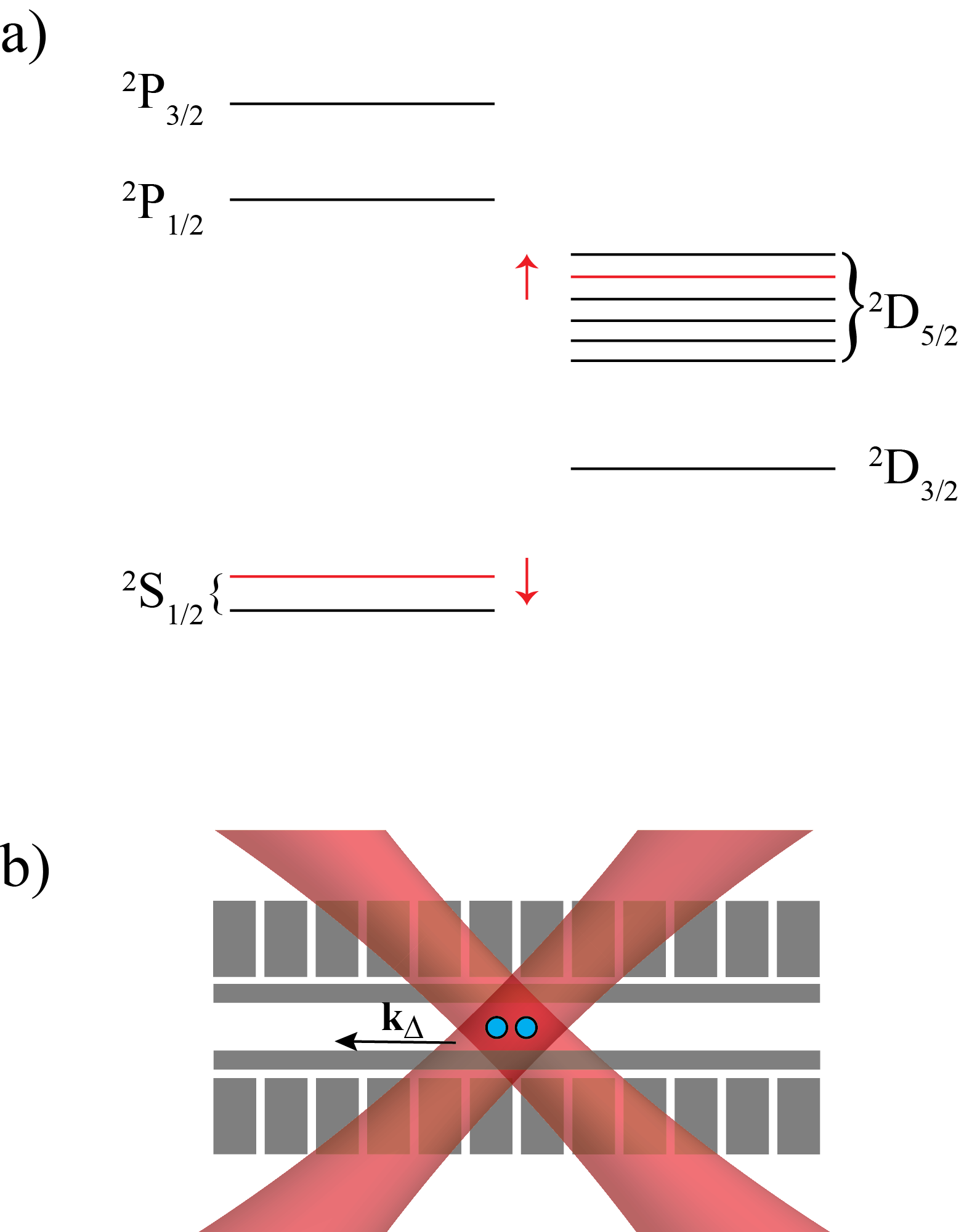}%
	\caption{\label{fig:TrapPic} (Color online) (a) Schematic level structure for the iso-electronic species $^{40}$Ca$^+$, $^{88}$Sr$^+$, $^{138}$Ba$^+$, and $^{226}$Ra$^+$ showing low-lying electronic states and the specific $m_J$ qubit levels (red) assumed for this work. (b) Example illustration of a $\sigma^z \sigma^z$ entangling gate experiment showing entangling laser beams intersecting at 90$^{\circ}$ (red) with linear polarizations out of the page, trap electrodes (gray), and difference-wavevector $\mathbf{k}_{\Delta}$ along the trap axis. Here we assume that the two-qubit gate is performed using the axial modes of a pair of identical or disparate co-trapped ions (blue circles). More extreme laser beam geometries (counter-propagating) are feasible at long (i.e. infrared) wavelengths.}
\end{figure}   

Two-qubit gates between trapped-ion qubits take advantage of the ions’ motional degrees of freedom to imprint quantum phases onto the system in a spin-dependent way. In the MS gate \cite{sorensen_entanglement_2000}, a laser beam with two tones is tuned near the upper and lower motional sidebands of the ions’ qubit transition frequency. This induces a force onto the ions whose strength depends on their states in the $\sigma^x\sigma^x$ or $\sigma^y \sigma^y$ basis. With appropriate detunings from the motional sidebands the system acquires the desired quantum geometric phase. The MS gate is often convenient, because it can be implemented with the same set of laser beams used to achieve one-qubit interactions (often at visible or IR wavelengths). Furthermore, it is compatible with qubits based on clock transitions with long coherence times \cite{langer_long-lived_2005}. Unfortunately, especially for optical-transition qubits, the MS scheme suffers from the presence of unwanted AC Stark shifts during the laser interaction which can fluctuate with small changes in laser intensity and result in $\sigma^z$ errors that do not commute with the two-qubit gate operation \cite{benhelm_towards_2008}. Moreover, the speed of the MS gate here varies only as the square root of laser intensity, making faster gates impractical and increasingly prone to errors from AC Stark shifts. Drifts in magnetic field or in qubit laser frequency also lead to (non-commuting) $\sigma^z$ errors in these systems. Calibrations for these drifts are time-intensive and may be difficult to perform quickly enough for accurate compensation of their effects.

The LS gate, thus far only experimentally demonstrated on hyperfine or Zeeman qubits, operates in a related manner but in a different basis \cite{leibfried_experimental_2003}. Here two laser beams, each far detuned from any transition, intersect to form an optical lattice at the ions’ positions. The beams do not directly induce qubit transitions, but their presence shifts the internal energy of the ions via the AC Stark effect. Because the size of this shift depends on the internal spin state, and because the ions reside in an optical lattice (with its accompanying optical intensity or polarization gradient), there is a force induced whose strength depends on the ions’ states in the $\sigma^z\sigma^z$ basis. With appropriate detunings from the motional sidebands the system acquires the desired quantum geometric phase. Because the LS gate uses a $\sigma^z\sigma^z$ interaction, it may be paired with straightforward dynamical decoupling pulses \cite{leibfried_experimental_2003,ballance_high-fidelity_2016} (a Hahn spin-echo pulse, for example) to eliminate $\sigma^z$ errors caused by unwanted AC Stark shifts or drifts in qubit frequency. Furthermore, the speed of the gate varies linearly with laser intensity, so that faster gates can be achieved. Unfortunately, because of the relatively small qubit splitting (MHz to GHz) in typical Zeeman and hyperfine systems, the laser beams cannot be sufficiently detuned from nearby optical transitions to suppress photon scattering errors without greatly suppressing the gate interaction strength \cite{ballance_high-fidelity_2016}. Furthermore, for these types of qubits the desired gate laser wavelengths typically lie within the UV range, a serious disadvantage from the perspectives of optical power generation, compatibility with optical fibers \cite{colombe_single-mode_2014} and waveguides \cite{niffenegger_integrated_2020}, and the accumulation of stray charges on trap surfaces \cite{wang_laser-induced_2011,allcock_heating_2012}.

Before 2015, the highest fidelity two-qubit gate had been performed with an MS interaction on an “optical” qubit in \Ca, with qubit levels separated by an $S-D$ transition wavelength \cite{benhelm_towards_2008}. In comparison with hyperfine and Zeeman qubits, an optical qubit benefits from higher detection fidelities \cite{myerson_high-fidelity_2008}. Ions with low-lying metastable levels also have the advantage of several auxiliary states which can be used to “hide” one ion from decoherence incurred while detecting the state of another nearby ion \cite{haffner_quantum_2008}. The wavelengths for manipulation of an optical qubit typically lie in the visible domain making the light easier to generate, easier to transport in fibers and waveguides, and less prone to creating stray charge in ion traps than UV light. Unfortunately, an optical qubit generally suffers more from decoherence caused by magnetic-field variation and beam intensity fluctuations, especially during MS two-qubit gate operations. Mølmer-Sørensen gates on optical qubits also impose stringent requirements on gate laser phase and frequency stability, where stabilization techniques are usually adapted directly from state-of-the-art optical clock experiments \cite{benhelm_ca$mathplus$quantum_2009}.

To date, a LS gate of the type described above has not been implemented in optical-transition qubits such as the long-lived $S_{1/2}-D_{5/2}$ superpositions in \Ca, \Sr, \Ba, and \Ra. Here we propose such an OTDF gate acting on the $S_{1/2}-D_{5/2}$ transitions in these species. Our restriction to nuclear-spin-free isotopes throughout this Article is for the sake of simplicity; extension to ions with hyperfine structure --- even including optical-clock-transition qubits~\cite{kim_geometric_2008} --- is straightforward. Expanding on recent dual-species entanglement work~\cite{ballance_hybrid_2015,hughes_benchmarking_2020}, the OTDF gate naturally operates without modification on pairs of the heterogeneous alkaline-earth ion species Ca$^+$, Sr$^+$, Ba$^+$, and Ra$^+$ in superpositions of their respective $S$ and $D$ levels. Furthermore, due to the broad wavelength tunability of the OTDF gate, the lighter species Be$^+$ and Mg$^+$ may each be paired with a heavier alkaline earth for dual-species entangling gates combining traditional hyperfine or Zeeman qubits (Be$^+$, Mg$^+$) with optical qubits (Ca$^+$, Sr$^+$, Ba$^+$, Ra$^+$). 

The OTDF gate can leverage large laser detunings (because of the $>100$ THz qubit splitting) to suppress photon scattering errors and can be implemented over a wide range of laser wavelengths even into the near- and short-wavelength infrared. Beyond their technological advantages, longer laser wavelengths yield smaller Lamb-Dicke factors, thereby relaxing ion temperature and trap confinement requirements. 

The remainder of this Article is structured as follows. Sec.~\ref{sec:GateOverview} provides a general description of the OTDF gate operations. Sec.~\ref{sec:ACSS} summarizes calculation of the AC Stark shifts and photon scattering rates for the four species under consideration. Sec.~\ref{sec:SameSpecies} presents gate durations and intrinsic error rates for identical ion pairs based on the calculations of Sec.~\ref{sec:ACSS}, and Sec.~\ref{sec:MultiSpecies} presents the gate durations and error rates for heterogeneous pairs. The full derivations for AC Stark shifts, photon scattering rates, gate durations, and intrinsic error rates are presented in the three appendices.

\section{Gate Overview} \label{sec:GateOverview}
The basic theory of LS gate operation has been well explored in the literature \cite{leibfried_experimental_2003,ballance_hybrid_2015,ballance_high-fidelity_2016}. For completeness, we provide a detailed derivation of OTDF gate operation in a multispecies system in Appendix~\ref{appendixA}. To perform entangling gates with axial modes of motion, we consider two optical-dipole-force (ODF) beams at angular frequencies $\omega_L$ and $\omega_L+2\omega_\Delta$ propagating $90^{\circ}$ to one another to form a traveling optical lattice overlapping a two-ion crystal with a motional mode frequency $\Omega_k$ (see Fig.~\ref{fig:TrapPic}(b)). At any instant in time, the beams form an optical lattice with wavelength $\lambda_\mathrm{eff}=2\sqrt{2} \pi c/\omega_L$ along the ion trap symmetry axis, where $c$ is the speed of light.

Because the two beams are detuned from each other, the optical lattice moves through a distance $\lambda_\mathrm{eff}$ in the interval $2\pi/(2\omega_\Delta)$, thereby modulating the optical dipole force that each ion experiences. For a constant detuning $\delta_k$ from the gate mode $2\omega_\Delta\equiv\Omega_M+\delta_k$, the mode is first excited and then de-excited and follows a circle in position-momentum (x-p) phase space. After a duration $2\pi/\delta_k$ the motion has returned to its initial phase-space location, but the quantum state has acquired an overall geometric phase proportional to the area enclosed by the path in phase space.

The ODF strength depends on the internal state ($\qubitup$ or $\qubitdown$) of each ion, so that the geometric phase is state-dependent: the four possible states $\{\ket{\uparrow\uparrow},\ket{\uparrow\downarrow},\ket{\downarrow\uparrow},\ket{\downarrow\downarrow}\}$ acquire phases $\{\Phi_{\uparrow\uparrow},\Phi_{\uparrow\downarrow},\Phi_{\downarrow\uparrow},\Phi_{\downarrow\downarrow}\}$. To symmetrize the interaction we imagine performing two such pulses separated by a spin echo pulse such that the total acquired phases become $\{\Phi_{\uparrow\uparrow\downarrow\downarrow},\Phi_{\uparrow\downarrow\downarrow\uparrow},\Phi_{\uparrow\downarrow\downarrow\uparrow},\Phi_{\uparrow\uparrow\downarrow\downarrow}\}$. For an appropriate choice of ODF strength, the ideal two-qubit phase gate ($\Phi_{\uparrow\downarrow\downarrow\uparrow}=\Phi_{\uparrow\uparrow\downarrow\downarrow}\pm\frac{\pi}{2}$) is realized~\cite{leibfried_experimental_2003}. This symmetrization enables the use of ions of disparate species, qubit states with vastly differing AC Stark shifts, and ions within an ODF intensity gradient (among other benefits; see Appendix~\ref{appendixA} for more details).

Because previous LS gate implementations used hyperfine or Zeeman qubits and gate laser wavelengths tuned near the $S_{1/2}-P_{1/2}$ and $S_{1/2}-P_{3/2}$ transitions, these experiments could leverage ODF laser polarizations which partially or completely nulled the differential AC Stark shift between $\qubitup$ and $\qubitdown$ states while still achieving finite optical-dipole forces via a non-vanishing polarization gradient \cite{leibfried_experimental_2003,baldwin_high_2020}. In our proposed scheme, the ODF beams may be detuned far ($>100$ THz) from any $S-P$ or $D-P$ transitions, and it is not possible to eliminate the differential Stark shift by any choice of polarization. Instead, one can choose an arbitrary polarization (speed is maximized when the two beam polarizations are matched: each linear and orthogonal to the laser-beam k-vectors) and rely on an optical intensity gradient rather than on a polarization gradient to create the necessary ODF. However, the resulting differential Stark shift induces unwanted $\sigma^z$ rotations onto the ions beyond the desired phase-gate interactions. In principle these could be accounted for with careful control and calibration of laser intensity, but in practice the use of a spin echo ensures that unwanted rotations from the first gate pulse will be canceled by analogous rotations from the second. The dependence of these unwanted rotations on the absolute optical phase of the lattice can be eliminated by turning the laser pulses on and off adiabatically~\footnote{The accumulated qubit phase after a gate pulse may be expressed as the Fourier transform of the pulse envelope at the optical beatnote frequency. Adiabatic ramps lead to negligible Fourier components at a given beatnote frequency.}.

\section{AC Stark Shifts and Scatter Rates} \label{sec:ACSS}
The speed of the OTDF gate outlined above is governed by the magnitude of the AC Stark shift induced by the gate beams, which is dominated by $D-P$ and $S-P$ electric dipole transitions. The minimum entangling gate duration for two phase-space loops (derived in Appendix~\ref{appendixA} and reproduced here) is 
\begin{equation}
\label{eqn:tau_g}
    \tau_g=\sqrt{\frac{2\pi^2}{\left|\eta_{k1} \eta_{k2} \cos(\phi_{0,2}-\phi_{0,1}) \Delta_{0,\Delta}(x_{0,1}) \Delta_{0,\Delta}(x_{0,2})\right|}}.
\end{equation}
In addition to its dependence on the effective Lamb-Dicke parameters $\eta_{kj}$ and ion spacing (captured via $\phi_{0,j}$), the duration is inversely proportional to the geometric mean of the beams' differential Stark shifts at the two ion locations, $\Delta_{0,\Delta}(x_{0,j})$.

\begin{figure*}
	\includegraphics{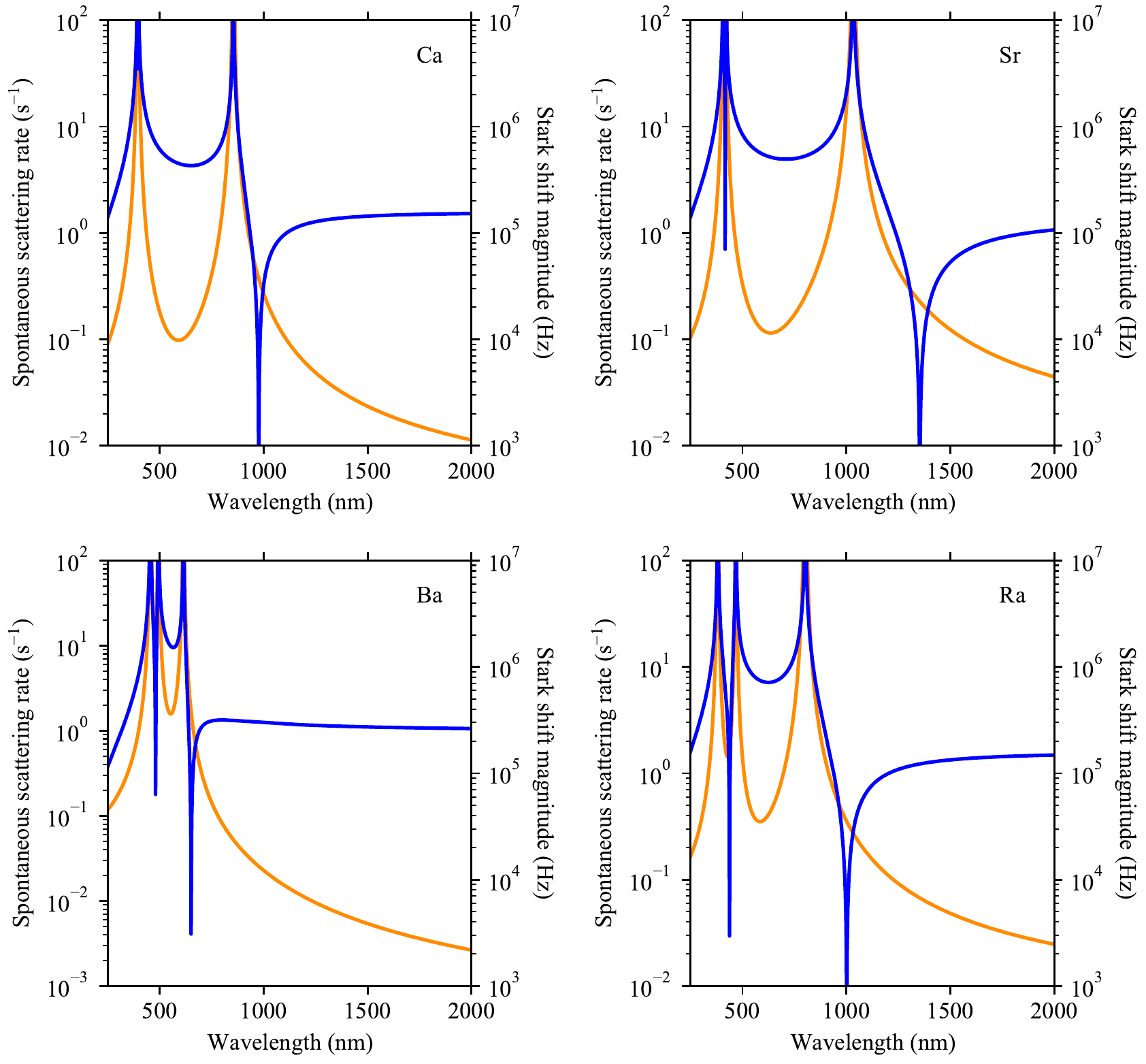}
	\caption{\label{fig:ACSS} (Color online) Spontaneous photon scattering decoherence rate (orange line, left axis) and AC Stark shift magnitude (blue line, right axis) versus laser wavelength for (a) \Ca, (b) \Sr, (c) \Ba, and (d) \Ra. All curves shown assume a single ion at the center of a single laser beam with a total power of 100~mW, a Gaussian waist of 10~$\mu$m (peak intensity 637~MW/m$^2$), and linear (equal $\sigma^+$ and $\sigma^-$) polarization. A gate would generally be operated far enough from the resonant wavelengths to reduce spontaneous scattering to an acceptable level.} 
\end{figure*}

Two unavoidable sources of error are spontaneous photon scattering and metastable-state decay. A complete theoretical description of the former, including both elastic and inelastic terms, is presented in Appendix~\ref{appendixB}. Summarizing here, the total intrinsic two-ion decoherence rate, $\Gamma^{(2)}_{\mathrm{tot}}$, is
\begin{equation}
\label{eqn:gamma}
    \Gamma_{\mathrm{tot}}^{(2)} = \sum_{j=1}^2 \left( \Gamma_{\mathrm{el},j} + \Gamma_{\mathrm{in},j} + \frac{1}{2} A_{D,j} \right),
\end{equation}
where $\Gamma_{\mathrm{el},j} ~ \left(\Gamma_{\mathrm{in},j}\right)$ is the elastic (inelastic) photon scattering decoherence rate of ion $j$ due to the OTDF gate laser beams as defined in Appendix~\ref{appendixB}. The final term in Eq.~\ref{eqn:gamma} describes spontaneous decay of the metastable superposition of $S$ and $D$ levels, which is one-half of the $D$-state decay rate of ion $j$. Given the finite lifetime of the $D$ levels, it might be favorable to adopt a `hybrid' qubit encoding whereby population is cycled between Zeeman qubits for storage and optical qubits for two-qubit interactions as recently discussed in Ref.~\cite{mehta_integrated_2020}. We exclude Yb$^+$ from our OTDF gate analysis because of the relatively short lifetimes $(\lesssim60~\text{ms})$ of its metastable states~\cite{taylor_investigation_1997,yu_lifetime_2000}.

Reference~\cite{uys_decoherence_2010} describes the process by which \textit{elastic} photon scattering causes decoherence of a superposition state. In summary, if the elastic photon scattering amplitudes are different between the $\qubitup$ and $\qubitdown$ states, then the elastic scattering process will cause dephasing of the superposition at a rate proportional to the square of the difference between the scattering \textit{amplitudes} from $\qubitup$ and $\qubitdown$ (see Eq.~\ref{eqn:GammaEl}). In practice, qubit states with a frequency difference comparable to or larger than the gate laser beam detuning from the nearest intermediate state will suffer from non-negligible elastic photon scattering decoherence.   

Both the AC Stark shift and the scattering decoherence rate for a single \Ca, \Sr, \Ba, and \Ra ion are depicted in Fig.~\ref{fig:ACSS} over the wavelength range $300-2000$~nm at a single-laser-beam intensity of 637~MW/m$^2$ (e.g. 100~mW in a 10~$\mu$m Gaussian beam waist). We assume linear (equal $\sigma^+$ and $\sigma^-$) polarization, although polarization effects are negligible for the large detunings considered here. The narrow electric quadrupole resonances are not visible in the curves of Fig.~\ref{fig:ACSS}. We neglect their contribution to the AC Stark shifts and scattering rates, assuming that the OTDF gate laser frequencies are not directly resonant with any $S-D$ transitions~\footnote{Given the extremely large ratio of $A$-coefficient values between $P$ and $D$ levels of $\sim10^8$ for all species considered here, a gate laser detuning of $>100~\text{MHz}$ from the nearest $S - D$ transition is sufficient to neglect this effect for all species.}  
 
\section{Same Species Entanglement} \label{sec:SameSpecies}
\begin{figure*}
	\includegraphics{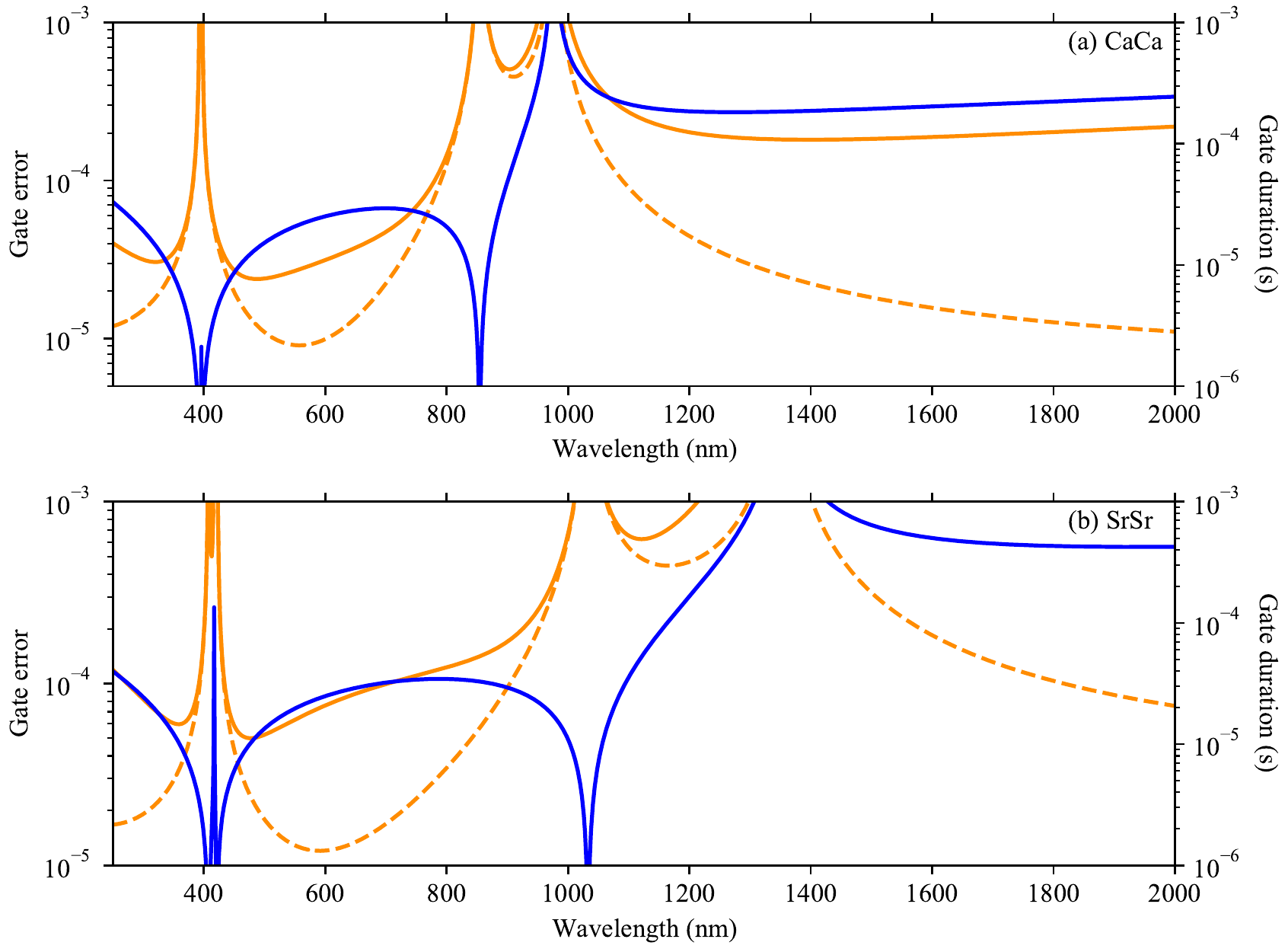}%
	\caption{\label{fig:SameSpec} (Color online) Total fundamental error  (solid orange line, left axis), photon scattering error (dotted orange line, left axis) and gate duration (blue line, right axis) for an OTDF two-qubit entangling gate between identical, co-trapped ion species (a) $^{40}\text{Ca}^+-^{40}\text{Ca}^+$ and (b) $^{88}\text{Sr}^+-^{88}\text{Sr}^+$. Total error includes both elastic and inelastic photon scattering decoherence as well as metastable decay. All curves assume a two-ion crystal confined in a potential that would yield a 2~MHz axial frequency for a single $^{40}\text{Ca}^+$ ion and centered upon two laser beams, each with a power of 100~mW, a 10~$\mu$m Gaussian waist (peak intensity 637~MW/m$^2$), an angle of 45$^\circ$ to the crystal axis, and linear (equal $\sigma^+$ and $\sigma^-$) polarization.}
\end{figure*}

\begin{figure*}
	\includegraphics{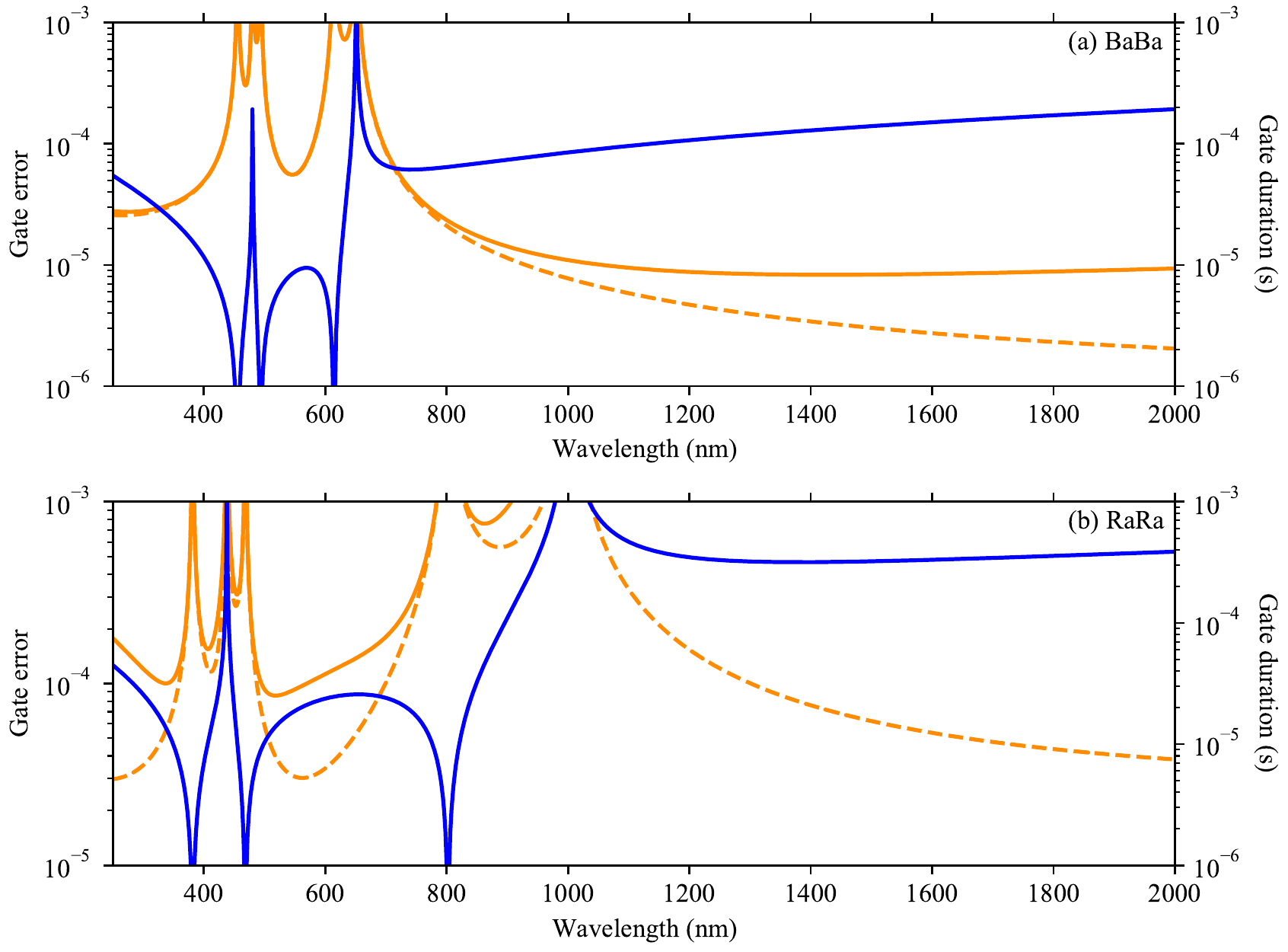}%
	\caption{\label{fig:SameSpec2} (Color online) Total fundamental error  (solid orange line, left axis), photon scattering error (dotted orange line, left axis) and gate duration (blue line, right axis) for an OTDF two-qubit entangling gate between identical, co-trapped ion species (a) $^{138}\text{Ba}^+-^{138}\text{Ba}^+$ and (b) $^{226}\text{Ra}^+-^{226}\text{Ra}^+$. Assumptions are the same as in Fig.~\ref{fig:SameSpec}.}
\end{figure*}

Given Eqs.~\ref{eqn:tau_g} and ~\ref{eqn:gamma}, it is straightforward to calculate the gate durations, $\tau_g$, and intrinsic errors, $\epsilon=\Gamma_{\mathrm{tot}}^{(2)}\tau_g$, for various ion pairs. We begin with a comparison of pairs consisting of the same ion species ($^{40}$Ca$^+$, $^{88}$Sr$^+$, $^{138}$Ba$^+$, and $^{226}$Ra$^+$) in Figs.~\ref{fig:SameSpec} and \ref{fig:SameSpec2}.
 Here, we assume the pair is trapped in a harmonic electrostatic potential providing a 2~MHz axial frequency to a single $^{40}$Ca$^+$ ion (we neglect the small changes in potential that would be required to achieve ion spacing optimized for gate speed). The chosen qubit states (Fig.~\ref{fig:TrapPic}(a)) are $\ket{^2S_{1/2}(m_J=1/2)}$ and $\ket{^2D_{5/2}(m_J=+3/2)}$. The ion pair is illuminated by two beams each with a power of 100~mW, a 10~$\mu$m Gaussian waist (peak intensity 637~MW/m$^2$ per beam), an angle of 45$^\circ$ to the crystal axis, and linear (equal $\sigma^+$ and $\sigma^-$) polarization as in Fig.~\ref{fig:TrapPic}(b).
 
 Note that at a given wavelength, the photon scattering decoherence rate (Eq.~\ref{eqn:Gamma_if}) is proportional to the laser intensity while the gate duration (Eq.~\ref{eqn:tau_g}) is inversely proportional to intensity. For a gate duration with a negligible contribution of $D$-state decay error, the fundamental two-qubit-gate error $\left( \Gamma_{\mathrm{tot}}^{(2)}\tau_g \right)$ is independent of laser intensity. Near the $S-P$ and $D-P$ resonant wavelengths the gate speeds grow quickly, but the photon scattering error rates increase faster still, so that wavelengths optimized for minimum error occur at relatively large detunings from these resonances. Similarly, the ``magic wavelengths" where the differential AC Stark shift approaches zero must also be avoided. 
 
 All four ions considered in this Article offer two qubit gate wavelengths in the visible range where the error is below $10^{-4}$ and the gate duration is a few 10's of $\mu$s. Calcium presents the lowest error in this range ($2.4\times10^{-5}$) while $^{138}$Ba$^+$ gates are particularly fast with durations $<10$~$\mu$s throughout the visible spectrum. Near the 1550-nm telecom wavelength $^{40}$Ca$^+$ attains a gate error of $1.9\times10^{-4}$ and $^{138}$Ba$^+$ achieves $<10^{-5}$ (the other two species have error $>10^{-3}$ so appear less useful in this range). Although the gate speeds here are slower than in the visible by an order of magnitude (due to a combination of smaller Lamb-Dicke parameters and smaller Stark shifts) and the gate errors (overwhelmed here by metastable decay) are correspondingly higher, the possibility to generate and modulate higher levels of infrared power mitigates these deficiences to some extent. The use of a phase-modulated retro-reflected beam geometry could recycle the incoming laser power and would allow in-situ lattice phase characterization and stabilization with high bandwidth, experimentally removing optical lattice phase fluctuations that can limit precision measurement experiments with trapped ions~\cite{gilmore_amplitude_2017}. 
 
 Despite the larger gate errors for all species in the infrared regime, quantum simulation experiments (e.g. variational quantum eigensolvers~\cite{hempel_quantum_2018,Wang_resource_optimized_2020}, quantum spin model simulations~\cite{britton_engineered_2012,jurcevic_direct_2017,zhang_observation_2017}, and quantum approximate optimization~\cite{pagano_quantum_2020}) may benefit from the more mature optical technologies at longer wavelengths. Furthermore, the larger Lamb-Dicke parameters at infrared gate laser wavelengths permit weaker trap confinement or imperfect ground-state cooling. Efficient ground state cooling of the relevant motional modes of many-ion crystals is particularly challenging, a problem which further motivates entangling gate schemes utilizing infrared wavelengths in both radiofrequency and Penning ion trap quantum simulation experiments.
 
\section{Multispecies Entanglement} \label{sec:MultiSpecies}
\begin{figure*}
	\includegraphics{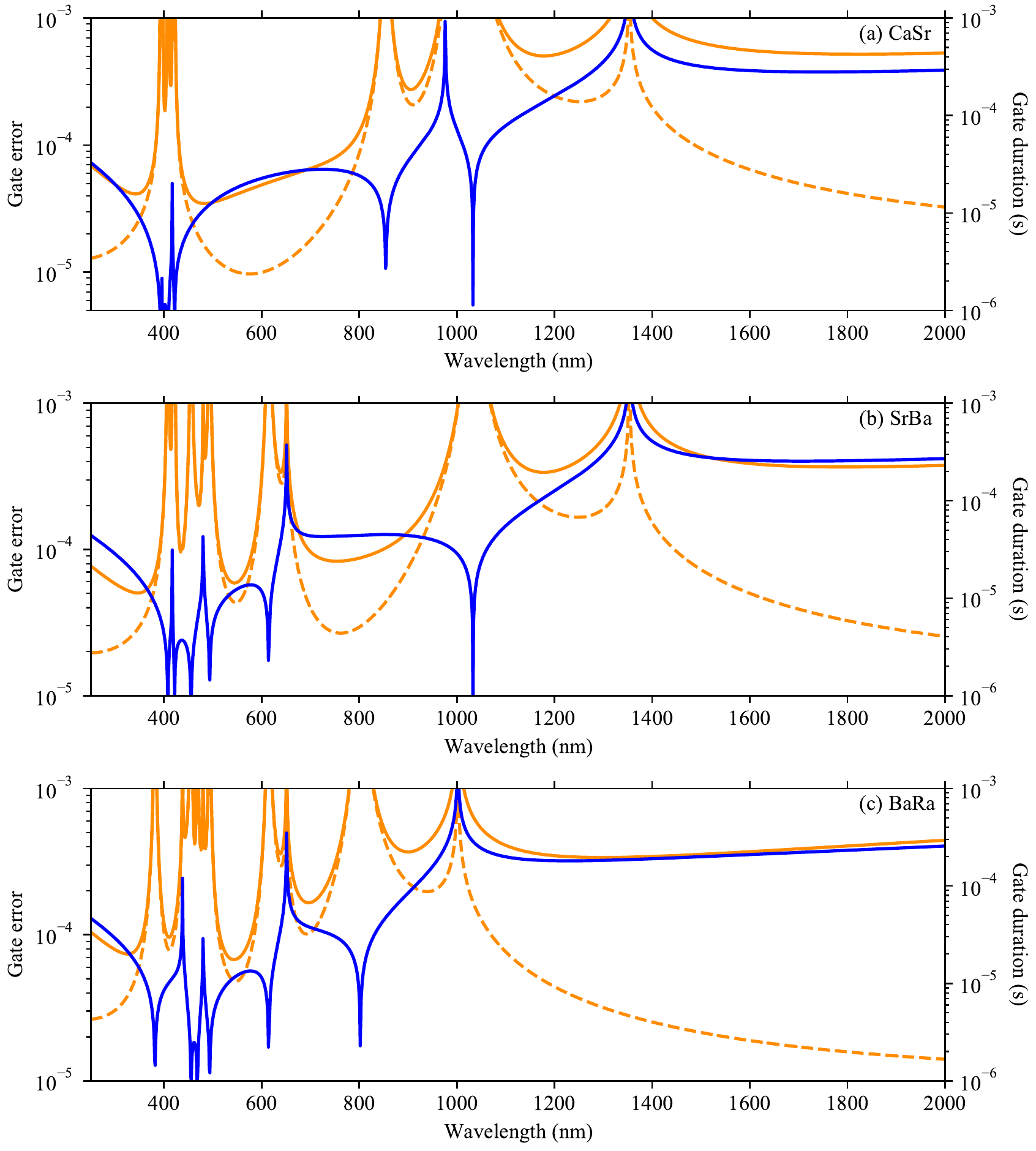}%
	\caption{\label{fig:DiffSpec} (Color online) Total fundamental error  (solid orange line, left axis), photon scattering error (dotted orange line, left axis) and gate duration (blue line, right axis) for an OTDF two-qubit entangling gate between pairs of heterogeneous-species, co-trapped ions (chosen for favorable mass ratios) (a) $^{40}\text{Ca}^+-^{88}\text{Sr}^+$, (b) $^{88}\text{Sr}^+-^{138}\text{Ba}^+$, and (c) $^{138}\text{Ba}^+-^{226}\text{Ra}^+$. Assumptions are the same as in Fig.~\ref{fig:SameSpec}.}
\end{figure*}

We performed similar calculations for pairs of heterogeneous ion species and present the results for ($^{40}$Ca$^+-^{88}$Sr$^+$, $^{88}$Sr$^+-^{138}$Ba$^+$, and $^{138}$Ba$^+-^{226}$Ra$^+$) in Fig.~\ref{fig:DiffSpec}. Other combinations yield similar gate errors and speeds but are not presented here due to the large difference in Mathieu stability parameter for ions with widely disparate masses~\cite{wineland_experimental_1997}. This challenge is not insurmountable but reduces the appeal of such mass combinations.
The trapping potential and laser beam characteristics are the same as for the homogeneous case (Section~\ref{sec:SameSpecies}). The dissimilar-ion behavior is more complex as a function of wavelength than the homogeneous case as there are twice as many resonant wavelengths that must be avoided. We reiterate that a spin flip between the two ODF pulses is essential to symmetrize the interaction here because of the mismatched ion Stark shifts. In general, the gate errors and speeds are comparable to the same-species gates over a wide range of wavelengths. We note that there are many regions where gate performance is comparable between a pair of identical ions and a pair of heterogeneous ions. For example, $\text{Ca}^+-\text{Sr}^+$ gates perform well in the range $\sim450-800$~nm which is also favorable for the $\text{Ca}^+-\text{Ca}^+$ combination. Accordingly, gates on both configurations could be achieved in a multi-ion system via the same OTDF laser beams.

The OTDF gate may also be extended to multispecies crystals that include $\text{Be}^+$ or $\text{Mg}^+$. This is possible since all of the alkaline earth species considered in Fig.~\ref{fig:ACSS} have favorable AC Stark shifts and photon scatter rates near the traditional LS gate wavelengths of $\sim 313~\text{nm}$ for $\text{Be}^+$~\cite{leibfried_experimental_2003} and $\sim 280~\text{nm}$ for $\text{Mg}^+$. Extreme mismatch of atomic masses remains a concern for co-trapped ions, therefore $\text{Be}^+$ and $\text{Mg}^+$ would be best paired with \Ca or \Sr.

\section{Extrinsic Gate Errors} \label{sec:GateErrors}
Beyond fundamental physics constraints, realistic technical limitations to our proposed two-qubit gate must also be considered. Some important technical noise sources include gate laser intensity instability, limitations to spin-echo dynamical decoupling, and trap frequency variations caused by dielectric charging or drift in the trapping potential electronics. We briefly discuss each of these potential error sources below.

Assuming the laser intensity remains fixed during the gate but varies from shot to shot with a fractional variation $\frac{\Delta I}{I}$ away from the optimum value $I$, to lowest order the associated error is
\begin{equation}
\epsilon = \frac{\pi^2}{4} \left(\frac{\Delta I}{I}\right)^2.
\end{equation}
An identical error is present in other experimentally demonstrated gates \cite{benhelm_towards_2008}.

A unique class of errors is that associated with the non-zero Stark shift required for OTDF operation. For a pulse intensity profile rectangular in time, this Stark shift varies sinusoidally about its mean. Because the OTDF gate incorporates a spin echo, it is not necessary to calibrate the precise phase rotation imprinted by this Stark shift, but it is necessary to ensure that the time integral of the second pulse's intensity matches that of the first. We consider the integral of the sinusoidal component separately from that of the quasi-DC component. The sinusoidal component can in principle be controlled via a tight phase lock between the lattice optical phase and the pulse envelope, although a simpler solution employs an adiabatic ramp up and down of the pulse so that the integral becomes negligibly small. The quasi-DC term leads to the requirement that the time-averaged intensity of the second pulse matches that of the first. We note that during a pulse the quasi-DC Stark term from each beam (assuming optimum ion spacing) leads to a phase
\begin{equation}
\begin{split}
\phi_0&=\Delta_{0,\Delta}\frac{\tau_g}{2}\\
&=\frac{\pi}{\sqrt{2\eta_{k1}\eta_{k2}}}
\end{split}
\end{equation}
imprinted onto each ion. A fractional intensity difference $\frac{\Delta I_{12}}{I}$ between the first and second pulses therefore leads to an uncompensated phase shift $\Delta\phi_0$ with associated error
\begin{equation}
\begin{split}
    \epsilon&=\Delta\phi_0^2 \\
    &=\left(\phi_0 \frac{\Delta I_{12}}{I}\right)^2\\
    &=\frac{\pi^2}{2\eta_{k1}\eta_{k2}}\left(\frac{\Delta I_{12}}{I}\right)^2.
\end{split}
\end{equation}
For a representative $\eta_{ki}\sim0.07$ this requires $\frac{\Delta I_{12}}{I}\lesssim 3\times10^{-4}$ to achieve $\epsilon<10^{-4}$. While this appears to be a stringent requirement in terms of absolute intensity stability, the demand here is only on the relative stability between first and second pulses, which is significantly more feasible~\cite{thom_intensity_2018}. In different terms, the gate is sensitive to the Allan deviation of the unwanted $\sigma^z$ rotation over the gate time scale, not its absolute value.

If either the gate mode frequency or the lattice traveling wave frequency changes from the intended value by $\Delta\omega$, then the detuning $\delta_k$ of the oscillating dipole force from the gate mode will change to $\delta_k+\Delta\omega$ leading to errors. This source of error is not unique to the OTDF gate; it can be reduced through appropriate Walsh modulation of the OTDF beam phase difference \cite{hayes_coherent_2012}, although not below the value (to lowest order)
\begin{equation}
\epsilon = \frac{\pi^2}{4} \left(\frac{\Delta \omega}{\delta_k}\right)^2.
\end{equation}

\section{Conclusions}
We have presented a detailed analysis of a two-qubit entangling gate based on optical-dipole forces acting on optical-transition qubits capable of an error rate below $10^{-4}$ over a wide range of wavelengths. An entangling gate with wavelength tunability over a sizeable fraction of the laser-accessible electromagnetic spectrum gives an unprecedented degree of flexibility with respect to, for example, Lamb-Dicke confinement, required laser power, and gate fidelity, parameters that may be traded in different experimental circumstances. 

The OTDF gate scheme is naturally suited for entangling ions of different species or for optical-clock-transition qubits (in contrast with some earlier LS gates). In systems where entangling gates between both similar and dissimilar pairs of ions is required, the same beams can be used for both gates. Future work could include extension of this OTDF gate technique to atom-molecule or molecule-molecule entanglement using metastable superpositions of molecular ion states, augmenting recent molecular ion gate proposals~\cite{hudson_dipolar_2018,campbell_dipole-phonon_2019} and demonstrations~\cite{lin_quantum_2020}. The incorporation of spin-echo sequences into the gate further improves the viability of magnetic-field-sensitive optical-transition qubits. We anticipate that this wavelength-insensitive OTDF gate will significantly expand the experimental toolbox available to ion trap quantum computing researchers.

\begin{acknowledgments}
We acknowledge funding from GTRI for preparation of this manuscript. We also thank David Hayes, Adam  Meier, and Nicholas Guise for useful discussions.
\end{acknowledgments}

\appendix
\section{AC Stark Shifts}
\label{appendix_stark_shifts}
The OTDF gate relies on a differential AC Stark shift between an $S$ and a $D$ electronic state. For a given state $\ket{i}$, the AC Stark shift induced by a laser beam with electric field amplitude $\mathcal{E}_0$ and polarization components $\hat{\epsilon}_q$ ($q={0,\pm 1}$) due to off-resonant dipole interaction with another state $\ket{k}$ is \cite{budker_atomic_2004}
\begin{equation}
\label{eqn:acss}
    \Delta E_k=\left(\frac{\mathcal{E}_0^2 \hat{\epsilon}_q^2 \mu_{ki,q}^2 }{4\hbar}\right)\left(\frac{1}{\omega_{ki}-\omega_L} + \frac{1}{\omega_{ki}+\omega_L}\right),
\end{equation}
where
\begin{equation}
    \mu_{ki,q}^2 \equiv |\bra{k} \mu_q \ket{i}|^2
\end{equation}
and $\mu_q$ are the components of the electric-dipole operator. Here $\omega_L$ is the angular frequency of the laser and $\omega_{ki}\equiv (\omega_k - \omega_i)$ is the transition angular frequency between states $i$ and $k$. From the Wigner-Eckart theorem, the matrix elements are
\begin{align}
    & |\bra{k}\mu_q\ket{i}|^2 = \nonumber \\ 
    &\frac{3\pi \epsilon_0 \hbar c^3}{\omega_{ki}^3} A_{J_k J_i} (2 J_k+1)
    \begin{pmatrix}
    J_i & 1 & J_k \\
    m_i & q & -m_k
    \end{pmatrix}^2,
\end{align}
where $\hbar$ is the reduced Planck constant and $\epsilon_0$ is the permittivity of free space. The coefficient $A_{J_k J_i}$ is the spontaneous emission rate between an excited state $\ket{k}$ and the state $\ket{i}$, while the total electronic angular momentum of a state $\ket{a}$ and its projection along the quantization axis are represented by $J_a$ and $m_a$, respectively. For this work we assume that all excited, intermediate states $\ket{k}$ are sublevels of the $P_{1/2}$ and $P_{3/2}$ manifolds.

For small laser detunings the first term in Eq.~\ref{eqn:acss} dominates, but for larger detunings the second term (the Bloch-Siegert shift caused by the counter-rotating electric field) must also be included. Making the substitution $\mathcal{E}_0^2=2 I/(\epsilon_0 c)$, which relates the electric field amplitude to the intensity $I$, and summing over all intermediate states $\ket{k}$ gives the total energy shift:
\begin{equation}
    \Delta E=\sum_{k,q} \left\{\left(\frac{I \hat{\epsilon}_q^2 \mu_{ki,q}^2}{2\epsilon_0 \hbar c}\right)\left(\frac{1}{\omega_{ki}-\omega_L} + \frac{1}{\omega_{ki}+\omega_L}\right) \right\}.
\end{equation}

\section{Intrinsic Decoherence due to Spontaneous Decay and Photon Scattering}
\label{appendixB}
The OTDF entangling gate is performed on a metastable superposition of $S$ and $D$ electronic states and is subject to two intrinsic sources of decoherence: off-resonant photon scattering from the gate laser beams and spontaneous decay from $D\rightarrow S$.  The $S-D$ transitions of \Ca, \Sr, \Ba, and \Ra are at optical wavelengths spanning from to 674 nm to 2050 nm. The OTDF gate laser beams will induce off-resonant spontaneous photon scattering events via the excited $P_{1/2}$ and $P_{3/2}$ levels, which will cause qubit decoherence during the entangling operation. With such a large qubit splitting relative to gate laser beam detuning from the $P$ manifold, both \textit{elastic} and \textit{inelastic} photon scattering decoherence must be considered.

To treat inelastic (Raman) scattering, we begin with a general expression for the differential photon scattering cross section, $d\sigma/d\Omega$, from Ref.~\cite{rodney_loudon_quantum_2000}:
\begin{widetext}
\begin{equation} \label{diffscatter}
    \frac{d\sigma}{d\Omega}^{i\rightarrow f} =  \frac{\omega_L(\omega_L-\omega_{fi})^3}{(4\pi\epsilon_0)^2\hbar^2 c^4}
    \left | \sum_k \left( \frac{\bra{f}\hat{\epsilon}_s \cdot \vec{\mu} \ket{k}\bra{k}\hat{\epsilon}\cdot\vec{\mu}\ket{i}}{\omega_{ki}-\omega_L} + \frac{\bra{f}\hat{\epsilon} \cdot \vec{\mu} \ket{k}\bra{k}\hat{\epsilon}_s\cdot\vec{\mu}\ket{i}}{\omega_L+\omega_{kf}}
    \right) \right |^2 .
\end{equation}
\end{widetext}
In the above expression, $\vec{\mu}$ is the electric dipole operator. As in Appendix~\ref{appendix_stark_shifts}, transition angular frequencies between states $a$ and $b$ are written as $\omega_{ba}\equiv (\omega_b-\omega_a)$. Photon scattering occurs between an initial state $\ket{i}$ and  final state $\ket{f}$ via intermediate excited (i.e. $P$) atomic states $\ket{k}$. The gate laser polarization unit vector is $\hat{\epsilon}$ and the polarization of the scattered photon is $\hat{\epsilon}_s$. The first (`difference-frequency') term in the sum of Eq.~\ref{diffscatter} corresponds to off-resonant scattering through state $\ket{k}$, with the spontaneously-emitted photon leaving the atom in state $\ket{f}$. The second (`sum-frequency') term describes the opposite process, whereby a spontaneously-emitted photon couples state $\ket{i}$ to $\ket{k}$ and the gate laser photon connects $\ket{k}$ to $\ket{f}$. The `sum-frequency' term may be neglected in experiments where $\omega_{ki}-\omega_L$ is much smaller (for some $k$) than $\omega_L+\omega_{kf}$. However, this term is non-negligible in the far-detuned gate laser regimes presented in this Article.

We integrate Eq.~\ref{diffscatter} over all solid angle and multiply by the incoming photon flux $(I/(\hbar \omega_L))$ to obtain the total scatter rate from state $\ket{i}$ to $\ket{f}$ through all intermediate states $\ket{k}$:
\begin{widetext}
\begin{equation} \label{eqn:Gamma_if}
    \Gamma^{i \rightarrow f} = 
    \frac{3\pi c^2 I}{2\hbar} (\omega_L-\omega_{fi})^3
     \sum_{s=\{0,\pm1\}} \left | \sum_k (2J_k+1) \sqrt{ \frac{A_{J_k J_i} A_{J_k J_f}}{\omega_{ki}^3\omega_{kf}^3}}
     \left( \frac{\bra{f} T_s^1 \ket{k}\bra{k}\hat{\epsilon}\cdot\vec{T}\ket{i}}{\omega_{ki}-\omega_L} +
     \frac{\bra{f} \hat{\epsilon}\cdot\vec{T} \ket{k}\bra{k}T_s^1\ket{i}}{\omega_L+\omega_{kf}} 
     \right)
     \right |^2,
\end{equation}
\end{widetext}
where we have used the Wigner-Eckart theorem to replace the electric dipole operators with unitless spherical tensors, $T$. The coefficients $A_{J_kJ_i}$ and $A_{J_kJ_f}$ are the spontaneous emission rates between an intermediate level with total electronic angular momentum $J_k$ and an initial or final level with angular momentum $J_i$ and $J_f$, respectively. Here we assume that all intermediate states are sublevels of the $P_{1/2}$ and $P_{3/2}$ excited states and that the $S_{1/2}$, $D_{3/2}$, and $D_{5/2}$ sublevels can be initial or final states. In other words, only electric-dipole-allowed scattering channels are considered. Furthermore, note that a Raman scattering event can only occur if $\omega_L > \omega_{fi}$, otherwise the scattering rate is negative.

The matrix elements of Eq.~\ref{eqn:Gamma_if} may be evaluated as
\begin{align} \label{melement}
    & \bra{a}T^1_q\ket{b} = \\
    &(-1)^{J_a-m_a}(-1)^{L_a+S+J_b+1}
    \begin{pmatrix}
    J_a & 1 & J_b \\
    -m_a & q & m_b
    \end{pmatrix}, \nonumber
\end{align}
where $m_a$ is the projection of state $a$ with angular momentum $J_a$ on the external magnetic field axis, $L_a$ is the electronic orbital angular momentum of state $a$, and $S$ is the total electronic spin of the given level ($S=1/2$ for all states considered in this Article). The photon polarization state for electric dipole transitions is $q \in \{ -1,0,+1\}$. The rightmost element of Eq.~\ref{melement} is a Wigner-3j symbol. 

The total \emph{inelastic} scattering rate of a superposition of $\ket{\uparrow}$ and $\ket{\downarrow}$ states of a \textit{single} ion (assuming equal populations in each state) is
\begin{equation}
    \Gamma_{\mathrm{in}} = \frac{1}{2}\left( \sum_{f \neq \uparrow} \Gamma^{\uparrow\rightarrow f}
    + \sum_{f \neq \downarrow} \Gamma^{\downarrow\rightarrow f} \right).
\end{equation}

To obtain the \emph{elastic} scattering decoherence rate, we take the scattering amplitude from Eq.~\ref{eqn:Gamma_if} and apply the result of Ref.~\cite{uys_decoherence_2010}. For a superposition of $\ket{\uparrow}$ and $\ket{\downarrow}$ states of a single ion (assuming equal populations in each state),
\begin{equation} \label{eqn:GammaEl}
    \Gamma_{\mathrm{el}} = \frac{3\pi c^2 I}{4\hbar}\sum_{s=\{0,\pm1\}} \left( \chi_s^{\uparrow\rightarrow\uparrow} - \chi_s^{\downarrow\rightarrow\downarrow} \right)^2,
\end{equation}
where we define
\begin{widetext}
\begin{equation}
    \chi_s^{i\rightarrow i}\equiv \sum_k (2J_k+1) A_{J_kJ_i} \sqrt{\frac{\omega_L^3}{\omega_{ki}^6}} 
    \left( \frac{\bra{i} T_s^1 \ket{k}\bra{k}\hat{\epsilon}\cdot\vec{T}\ket{i}}{\omega_{ki}-\omega_L} +
     \frac{\bra{i} \hat{\epsilon}\cdot\vec{T} \ket{k}\bra{k}T_s^1\ket{i}}{\omega_L+\omega_{ki}}\right).
\end{equation}
\end{widetext}

Finally, the total intrinsic decoherence rate, $\Gamma_{\mathrm{tot}}^{(2)}$, of a two-ion system where both ions are in a superposition of their respective $S$ and $D$ states (with equal populations in each state) is
\begin{equation}
    \Gamma_{\mathrm{tot}}^{(2)} = \sum_j \left( \Gamma_{\mathrm{el},j} + \Gamma_{\mathrm{in},j} + \frac{1}{2} A_{D,j} \right),
\end{equation}
where $j$ is the ion index and $A_{D,j}$ is the spontaneous decay rate of the chosen $D$ level of ion $j$.

\section{Geometric Phase Gate Theory}
\label{appendixA}
The system Hamiltonian is given by
\begin{equation}
    H=H_m+H_s+H_{LI}.
\end{equation}
$H_m$ represents the ions' motional degrees of freedom, $H_s$ their spins, and $H_{LI}$ the laser-ion interactions which can couple spin to motion. In the following discussion we will make successive transformations to the spin and spin-motion interaction frames with Hamiltonians $H^s$ and $H^{s,m}$ in these frames:
\begin{equation}
    H^s=H_m+H^s_{LI},
\end{equation}
\begin{equation}
    H^{s,m}=H^{s,m}_{LI}.
\end{equation}
\subsection{Ion Motions from Mode Amplitudes}
Our first task is to determine effective Lamb-Dicke parameters for a multi-species, multi-ion, optical-lattice interaction. The following discussion was inspired by \cite{kielpinski_sympathetic_2000}; here we expand that treatment slightly. For a general configuration of $N$ ions in harmonic wells with linear couplings, we have the following Hamiltonian:
\begin{equation}
\label{eqn:general_hamiltonian}
    H_m=\frac{1}{2} \sum_i{m_i \Dot{x_i}^2} + \sum_{ij}{A_{ij} x_i x_j}.
\end{equation}
Here, each ion has mass $m_i$, and its displacement from equilibrium along the principal axis of its local potential is $x_i$ (there are three of these variables for each ion, one for each principal axis direction). We define
$X_i\equiv \sqrt{m_i} x_i$ (corresponding vector $\mathbf{X}$) and $B_{ij}=\frac{A_{ij}}{\sqrt{m_i m_j}}$ (corresponding matrix $\mathbf{B}$). Then
\begin{equation}
    H_m=\frac{1}{2} \sum_i{\Dot{X_i}^2} + \sum_{ij}{B_{ij} X_i X_j}.
\end{equation}
Let ${\mathbf{b}_k}$ be an orthonormal basis of eigenvectors of $\mathbf{B}$ with corresponding eigenvalues $\beta_k$, so that $\mathbf{B}\cdot\mathbf{b}_k=\beta_k \mathbf{b}_k$, and define components $q_k$ of $\mathbf{X}$ in this basis such that $\mathbf{X}=\sum_k{q_k \mathbf{b}_k}$. Then
\begin{equation}
\begin{split}
\sum_{ij}{B_{ij} X_i X_j}&=\sum_{ij}{B_{ij} X_j X_i} \\
&=(\mathbf{B}\cdot\mathbf{X})\cdot\mathbf{X} \\
&=\left(\sum_k{\mathbf{B}\cdot q_k \mathbf{b}_k}\right)\cdot \mathbf{X} \\
&=\left(\sum_k{\beta_k q_k \mathbf{b}_k}\right)\cdot \mathbf{X} \\
&=\sum_{kj}{\beta_k q_k \mathbf{b}_k \cdot q_j \mathbf{b}_j} \\
&=\sum_{k}{\beta_k q_k^2}.
\end{split}
\end{equation}
Therefore
\begin{equation}
    H_m=\frac{1}{2} \sum_i{\Dot{X_i}^2} + \sum_{k}{\beta_k q_k^2}.
\end{equation}
Now
$\sum_i{\Dot{X_i}^2}=\Dot{\mathbf{X}}\cdot\Dot{\mathbf{X}}=\sum_k{\Dot{q_k}^2}
$
, so
\begin{equation}
    H_m=\frac{1}{2} \sum_k{\Dot{q_k}^2 + \beta_k q_k^2}.
\end{equation}
This is the Hamiltonian for a collection of harmonic oscillator modes each with position $q_k$, with unit mass (in this choice of units), and with (angular) frequency $\Omega_k$ given by $\Omega_k^2=\beta_k$.

If we introduce raising and lowering operators $a_k^\dagger$ and $a_k$ for each mode in the collection in the usual way, then we can write the position operators
\begin{equation}
    q_k=\sqrt{\frac{\hbar}{2\Omega_k}}(a_k + a_k^\dagger).
\end{equation}
This expression gives us the transformation from normal-mode amplitudes to real-space coordinates:
\begin{equation}
\label{eqn:coordinate_transform}
\begin{split}
    x_i & =\frac{X_i}{\sqrt{m_i}}=\frac{1}{\sqrt{m_i}}\left(\sum_k{q_k \mathbf{b}_k}\right)_i \\
        & =\sum_k{b_{ki} \sqrt{\frac{\hbar}{2 m_i \Omega_k}}(a_k + a_k^\dagger).}
\end{split}
\end{equation}
Here we have defined the $i$th component of $\mathbf{b}_k$ as $(\mathbf{b}_k)_i\equiv b_{ki}$. What remains is to determine the matrix $\mathbf{B}$ and its eigenvectors for a particular configuration of ions.

Here we choose to consider motion only along the trap axis of linear symmetry, although our discussion could be expanded to encompass radial directions in a straightforward manner. We assume that ion $i$ by itself has frequency $\nu_i$  in the potential near equilibrium (in the absence of couplings), and we assume the equilibrium position of each ion along the axis is given by $R_i$, such that $R_j>R_i$ for $j>i$. Then
\begin{equation}
    H_m=\frac{1}{2} \sum_i{m_i \Dot{x_i}^2} + \frac{1}{2}\sum_i{m_i \nu_i^2 x_i^2} + V_\mathrm{eq}(\mathbf{x}) .
\end{equation}
Here $V_\mathrm{eq}(\mathbf{x})$ represents the potential between the ions at equilibrium, assumed to be second-order in the ion displacements. We assume that the first-order dependence has already been captured in the equilibrium positions of the ions.
To determine $V_\mathrm{eq}$, we expand the Coulomb interaction:
\begin{equation}
\begin{split}
    V_\mathrm{Coulomb}(\mathbf{x}) &=\left[\frac{e_0^2}{4\pi\epsilon_0}\sum_{i,j:j>i}{\frac{1}{R_j+x_j-R_i-x_i}}\right]\\
    &=\left[\frac{e_0^2}{4\pi\epsilon_0}\sum_{i,j:j>i}{\frac{1}{R_j-R_i}%
        \frac{1}{1-\frac{x_i-x_j}{R_j-R_i}}}\right]\\
    &=\Biggl[\frac{e_0^2}{4\pi\epsilon_0}\sum_{i,j:j>i}{\frac{1}{R_j-R_i}}\\
        &\times
\Biggl(1+\frac{x_i-x_j}{R_j-R_i}
        +\left(\frac{x_i-x_j}{R_j-R_i}\right)^2\Biggr)\Biggr].
\end{split}
\end{equation}
As discussed above, $V_\mathrm{eq}$ keeps only the second-order term:
\begin{equation}
\begin{split}
    V_\mathrm{eq}(\mathbf{x}) 
    &=\frac{e_0^2}{4\pi\epsilon_0}\sum_{i,j:j>i}{\frac{1}{|R_j-R_i|^3}%
        \left(x_i^2 - 2 x_i x_j + x_j^2\right)}\\
    &=\frac{1}{2}\frac{e_0^2}{4\pi\epsilon_0}\sum_{i,j:j \neq i}{\frac{1}{|R_j-R_i|^3}%
        \left(x_i^2 - 2 x_i x_j + x_j^2\right)}.
\end{split}
\end{equation}
To second order our Hamiltonian becomes
\begin{multline}
    H_m=\frac{1}{2} \sum_i{m_i \Dot{x_i}^2} + \frac{1}{2}\sum_i{m_i \nu_i^2 x_i^2}\\
    + \frac{1}{2}\frac{e_0^2}{4\pi\epsilon_0}\sum_{i,j:j\neq i}{\frac{1}{\left|R_j-R_i\right|^3}%
        \left(x_i^2 - 2 x_i x_j + x_j^2\right)}.
\end{multline}
Comparing this with Equation~\ref{eqn:general_hamiltonian} we determine that
\begin{equation}
    A_{ij}=\begin{cases}
    \frac{1}{2}m_i \nu_i^2 + \frac{e_0^2}{4\pi\epsilon_0}\sum_{k\neq i}{\frac{1}{\left|R_k-R_i\right|^3}}, & \text{if $i=j$}.\\
    \frac{-e_0^2}{4\pi\epsilon_0}\frac{1}{\left|R_j-R_i\right|^3}, & \text{if $i \neq j$}.
    \end{cases}.
\end{equation}

Now we specialize to the situation with only two ions, possibly with different masses. In this case
\begin{equation*}
    R_2-R_1=\left(\frac{e_0^2}{2\pi\epsilon_0 m_1 \nu_1^2}\right)^{1/3}=\left(\frac{e_0^2}{2\pi\epsilon_0 m_2 \nu_2^2}\right)^{1/3},
\end{equation*}
and we find
\begin{equation}
    A_{ij}=\begin{cases}
    m_i \nu_i^2, & \text{if $i=j$}.\\
    -\frac{1}{2}m_i \nu_i^2=-\frac{1}{2}m_j \nu_j^2, & \text{if $i \neq j$}.
    \end{cases}.
\end{equation}
In terms of the mass ratio $\mu=m_2/m_1$, the eigenvalues and eigenvectors of $\mathbf{B}$ are
\begin{equation}
\begin{split}
    \beta_1 &=\frac{\Omega_1^2}{2}=\frac{\left(1+\mu-\sqrt{1-\mu+\mu^2}\right)\nu_1^2}{2\mu},\\
    \beta_2 &=\frac{\Omega_2^2}{2}=\frac{\left(1+\mu+\sqrt{1-\mu+\mu^2}\right)\nu_1^2}{2\mu},\\
    \mathbf{b}_1 &= (b_{11},b_{12}) = \left(b_0,\sqrt{1-b_0^2}\right),\\
    \mathbf{b}_2 &= (b_{21},b_{22}) = \left(\sqrt{1-b_0^2},-b_0\right),\\
\end{split}
\end{equation}
where we have defined
\begin{equation}
    b_0 = \frac{1-\mu+\sqrt{1-\mu+\mu^2}}{\sqrt{\mu+\left(1-\mu+\sqrt{1-\mu+\mu^2}\right)^2}}.
\end{equation}
This gives all the information needed to determine real-space motions from normal-mode amplitudes via Equation~\ref{eqn:coordinate_transform}.
\subsection{Two-laser Stark Shifts}
Assume that the two optical-dipole-force beams are approximately plane-waves with frequencies $\omega_1$ and $\omega_2$ and wave-vectors $\mathbf{k}_1$ and $\mathbf{k}_2$. The laser-beam amplitudes are assumed constant over the extent of the ions' motions during the gate, but we allow for different intensities at the two ion positions. We assume that they have the same polarization, but the discussion can easily be broadened to arbitrary polarizations by considering each polarization component ($\pi$, $\sigma_+$, and $\sigma_-$) independently. In this more general case one would take the sum of the Stark shifts of the three polarization components to determine the overall optical-dipole force. The electric fields of the two beams along the polarization direction can be expressed as
\begin{equation}
\begin{split}
    E_1(\mathbf{r},t)= E_1(\mathbf{r}) \cos(\mathbf{k}_1\cdot\mathbf{r}-\omega_1 t - \phi_1),\\
    E_2(\mathbf{r},t)= E_2(\mathbf{r}) \cos(\mathbf{k}_2\cdot\mathbf{r}-\omega_2 t - \phi_2).
\end{split}
\end{equation}
Here $\mathbf{r}$ and $t$ represent the space and time coordinates of the fields. We define $E_0=E_1(\mathbf{r})$, $\Delta E=E_2(\mathbf{r})-E_1(\mathbf{r})$, so that the total electric field can be expressed as
\begin{widetext}
\begin{equation}
\begin{split}
    E(\mathbf{r},t)&=E_1(\mathbf{r},t) + E_2(\mathbf{r},t)\\ &=E_1(\mathbf{r}) \cos(\mathbf{k}_1\cdot\mathbf{r}-\omega_1 t - \phi_1)+ (E_2(\mathbf{r})-E_1(\mathbf{r})) \cos(\mathbf{k}_2\cdot\mathbf{r}-\omega_2 t - \phi_2)+ E_1(\mathbf{r}) \cos(\mathbf{k}_2\cdot\mathbf{r}-\omega_2 t - \phi_2) \\
    &=E_0 (\cos(\mathbf{k}_1\cdot\mathbf{r}-\omega_1 t - \phi_1) + \cos(\mathbf{k}_2\cdot\mathbf{r}-\omega_2 t - \phi_2))+{\Delta E}\cos(\mathbf{k}_2\cdot\mathbf{r}-\omega_2 t - \phi_2) \\
    &=2E_0 \cos(\frac{1}{2}[\mathbf{k}_1\cdot\mathbf{r}-\omega_1 t - \phi_1 + \mathbf{k}_2\cdot\mathbf{r}-\omega_2 t - \phi_2]) \cos(\frac{1}{2}[\mathbf{k}_1\cdot\mathbf{r}-\omega_1 t - \phi_1 - \mathbf{k}_2\cdot\mathbf{r}+\omega_2 t + \phi_2]])\\
    &\qquad +{\Delta E}\cos(\mathbf{k}_2\cdot\mathbf{r}-\omega_2 t - \phi_2)\\
    &=2E_0 \cos(\mathbf{k}_\Sigma\cdot\mathbf{r}-\omega_\Sigma t - \phi_\Sigma) \cos(\mathbf{k}_\Delta\cdot\mathbf{r}-\omega_\Delta t - \phi_\Delta) +{\Delta E}\cos(\mathbf{k}_2\cdot\mathbf{r}-\omega_2 t - \phi_2).
\end{split}
\end{equation}
\end{widetext}
We have defined $\mathbf{k}_\Sigma=\frac{1}{2}\left(\mathbf{k}_1+\mathbf{k}_2\right)$, $\mathbf{k}_\Delta=\frac{1}{2}\left(\mathbf{k}_1-\mathbf{k}_2\right)$, $\omega_\Sigma=\frac{1}{2}\left(\omega_1+\omega_2\right)$, $\omega_\Delta=\frac{1}{2}\left(\omega_1-\omega_2\right)$,
$\phi_\Sigma=\frac{1}{2}\left(\phi_1+\phi_2\right)$, and  $\phi_\Delta=\frac{1}{2}\left(\phi_1-\phi_2\right)$.
Here we are interested in the AC-Stark shift induced by these beams, so we take the square of the electric field and time-average over a duration long compared to optical frequencies but short compared to the frequency difference between the beams. The Stark shift is then
\begin{widetext}
\begin{equation}
\begin{split}
    \Delta(\mathbf{r},t)&=\gamma \langle E(\mathbf{r},t)^2 \rangle\\
    \Delta(\mathbf{r},t)&=\gamma \langle4E_0^2 \cos^2(\mathbf{k}_\Sigma\cdot\mathbf{r}-\omega_\Sigma t - \phi_\Sigma) \cos^2(\mathbf{k}_\Delta\cdot\mathbf{r}-\omega_\Delta t - \phi_\Delta)\\
    &\qquad + 4 E_0 \Delta E \cos(\mathbf{k}_\Sigma\cdot\mathbf{r}-\omega_\Sigma t - \phi_\Sigma) \cos(\mathbf{k}_\Delta\cdot\mathbf{r}-\omega_\Delta t - \phi_\Delta)  \cos(\mathbf{k}_2\cdot\mathbf{r}-\omega_2 t - \phi_2)\\
    &\qquad + {\Delta E}^2\cos^2(\mathbf{k}_2\cdot\mathbf{r}-\omega_2 t - \phi_2)\rangle\\
    &=\gamma (2E_0^2 \cos^2(\mathbf{k}_\Delta\cdot\mathbf{r}-\omega_\Delta t - \phi_\Delta)\\
    &\qquad+ \langle4 E_0 \Delta E \cos(\mathbf{k}_\Sigma\cdot\mathbf{r}-\omega_\Sigma t - \phi_\Sigma) \cos(\mathbf{k}_\Delta\cdot\mathbf{r}-\omega_\Delta t - \phi_\Delta) \cos(\mathbf{k}_2\cdot\mathbf{r}-\omega_2 t - \phi_2)\rangle\\
    &\qquad+\frac{1}{2}{\Delta E}^2)\\
    &=\gamma (2E_0^2 \cos^2(\mathbf{k}_\Delta\cdot\mathbf{r}-\omega_\Delta t - \phi_\Delta)\\
    &\qquad + \langle2 E_0 \Delta E \cos(\mathbf{k}_\Delta\cdot\mathbf{r}-\omega_\Delta t - \phi_\Delta) (\cos(\mathbf{k}_\Sigma\cdot\mathbf{r}-\omega_\Sigma t - \phi_\Sigma+\mathbf{k}_2\cdot\mathbf{r}-\omega_2 t - \phi_2)\\
    &\qquad + \cos(\mathbf{k}_\Sigma\cdot\mathbf{r}-\omega_\Sigma t - \phi_\Sigma-\mathbf{k}_2\cdot\mathbf{r}+\omega_2 t + \phi_2))\rangle +\frac{1}{2}{\Delta E}^2)\\
    &=\gamma (2E_0^2 \cos^2(\mathbf{k}_\Delta\cdot\mathbf{r}-\omega_\Delta t - \phi_\Delta)\\
    &\qquad+ \langle2 E_0 \Delta E \cos(\mathbf{k}_\Delta\cdot\mathbf{r}-\omega_\Delta t - \phi_\Delta) \cos(\mathbf{k}_\Delta\cdot\mathbf{r}-\omega_\Delta t - \phi_\Delta)\rangle
    +\frac{1}{2}{\Delta E}^2)\\
    &=\gamma [2(E_0^2+E_0 \Delta E) \cos^2(\mathbf{k}_\Delta\cdot\mathbf{r}-\omega_\Delta t - \phi_\Delta)+\frac{1}{2}{\Delta E}^2].
\end{split}
\end{equation}
\end{widetext}
$\gamma$ represents the constant of proportionality between intensity and Stark shift (see Eq.~\ref{eqn:acss}). Defining 
$\Delta_0(\mathbf{r})=\frac{\gamma}{2} (E_0^2+E_0\Delta E)=\sign(\gamma)\sqrt{(\frac{\gamma}{2}E_1^2)(\frac{\gamma}{2}E_2^2)}$ (the geometric mean of the individual beam Stark shifts) and $\Delta_0^\prime(\mathbf{r})=\frac{\gamma}{4} {\Delta E}^2+\Delta_0=\frac{1}{2}(\frac{\gamma}{2}E_1^2+\frac{\gamma}{2}E_2^2)$ (the arithmetic mean), we can further simplify this to
\begin{equation}
\label{eqn:stark_shift}
\begin{split}
    \Delta(\mathbf{r},t)&=\gamma [(E_0^2+E_0 \Delta E)\\ 
    &\quad \times \{1+\cos(2\mathbf{k}_\Delta\cdot\mathbf{r}-2\omega_\Delta t - 2\phi_\Delta)\}
    +\frac{1}{2}{\Delta E}^2]\\
    &=2\Delta_0(\mathbf{r})\cos(2\mathbf{k}_\Delta\cdot\mathbf{r}-2\omega_\Delta t - 2\phi_\Delta) + 2\Delta_0^\prime(\mathbf{r}).\\
\end{split}
\end{equation}

\subsection{Optical Dipole Force Hamiltonian}
The first term in Eq.~\ref{eqn:stark_shift} represents the desired optical-dipole potential. The second ($2\Delta_0^\prime(\mathbf{r})$) term represents additional Stark shifts. These must be accounted for during operation of the gate, either by the addition of appropriate phase rotations or through a spin-echo pulse. This term also leads to optical-dipole forces, but these are generally negligible unless there is significant spatial variation in beam intensity. Furthermore, for slowly time-varying laser pulses this term is off-resonant with any ion motion, further reducing its impact. We ignore it in the following discussion.

Each ion's possible states ($\ket{\uparrow}$ and $\ket{\downarrow}$) will experience a different Stark shift $\Delta_\uparrow(\mathbf{r},t)$ and $\Delta_\downarrow(\mathbf{r},t)$. In the spin interaction frame, the laser-ion Hamiltonian is given by
\begin{multline}
    H^s_{LI}=\sum_j \biggl(\frac{1}{2}(\Delta_\uparrow(\mathbf{r}_j,t)-\Delta_\downarrow(\mathbf{r}_j,t))\sigma_{z,j}\\
    +\frac{1}{2}(\Delta_\uparrow(\mathbf{r}_j,t)+\Delta_\downarrow(\mathbf{r}_j,t))\biggr),
\end{multline}
where the sum runs over the two ions. Defining the differential Stark shift $\Delta_\Delta=\Delta_\uparrow-\Delta_\downarrow$ and (twice) the common Stark shift $\Delta_\Sigma=\Delta_\uparrow+\Delta_\downarrow$ we find
\begin{equation}
\begin{split}
    H^s_{LI}&=\sum_j{\biggl(\frac{1}{2}\Delta_\Delta(\mathbf{r}_j,t)\sigma_{z,j}
    +\frac{1}{2}\Delta_\Sigma(\mathbf{r}_j,t)\biggr)}\\
    &=\sum_j \biggl( [\Delta_{0,\Delta}(\mathbf{r}_j)\sigma_{z,j}+\Delta_{0,\Sigma}(\mathbf{r}_j) ]\\
    &\quad \times \cos(2\mathbf{k}_\Delta\cdot\mathbf{r}_j-2\omega_\Delta t - 2\phi_\Delta)\biggr).
\end{split}
\end{equation}
Considering here motion only along the trap axis, we can write $\mathbf{k}_\Delta\cdot\mathbf{r}_j=k_\Delta (x_{0,j}+x_j)$ (where $x_{0,j}$ is the equilibrium position). We define
\begin{equation}
    \eta_{kj} = 2 k_\Delta b_{kj} \sqrt{\frac{\hbar}{2 m_j \Omega_k}}\\
\end{equation}
\begin{equation}
    \phi_{0,j} = 2 \phi_\Delta - 2 k_\Delta x_{0,j}
\end{equation}
and substitute Eq.~\ref{eqn:coordinate_transform} to find
\begin{equation}
\begin{split}
    H^s_{LI}&=\sum_j \biggl( [\Delta_{0,\Delta}(\mathbf{r}_j)\sigma_{z,j}+\Delta_{0,\Sigma}(\mathbf{r}_j) ]\\
    &\times \cos(2k_\Delta [x_{0,j}+x_j]-2\omega_\Delta t - 2\phi_\Delta)\biggr)\\
    &=\sum_j \biggl( [\Delta_{0,\Delta}(x_{0,j})\sigma_{z,j}+\Delta_{0,\Sigma}(x_{0,j}) ]\\
    &\times \frac{1}{2}\exp(i\{2 k_\Delta [x_{0,j}+x_j]-2\omega_\Delta t - 2\phi_\Delta\}) \biggr) + \mathrm{H.c.}\\
    &=\sum_j \frac{1}{2} \biggl( [\Delta_{0,\Delta}(x_{0,j})\sigma_{z,j}+\Delta_{0,\Sigma}(x_{0,j}) ]\\
    &\times \exp(i\{\sum_k [ \eta_{kj} (a_k+a_k^\dagger)]-2\omega_\Delta t - \phi_{0,j}\}) \biggr) + \mathrm{H.c.}\\
\end{split}
\end{equation}
Here we have assumed that the beam intensities do not vary appreciably over the extent of the ions' motions. In the Lamb-Dicke regime (keeping terms only to first order in $\{\eta_{kj}\}$)
\begin{equation}
\begin{split}
    H^s_{LI}&=\sum_j \frac{1}{2} \biggl( [\Delta_{0,\Delta}(x_{0,j})\sigma_{z,j}+\Delta_{0,\Sigma}(x_{0,j}) ]\\
    &\times \exp(i\{-2\omega_\Delta t - \phi_{0,j}\}) \\
    &\times \exp(i\{\sum_k [ \eta_{kj} (a_k+a_k^\dagger)]\})\biggr) + \mathrm{H.c.}\\
    &\approx \sum_j \frac{1}{2} \biggl( [\Delta_{0,\Delta}(x_{0,j})\sigma_{z,j}+\Delta_{0,\Sigma}(x_{0,j}) ]\\
    &\times \exp(i\{-2\omega_\Delta t - \phi_{0,j}\}) \\
    &\times (1+i\{\sum_k [ \eta_{kj} (a_k+a_k^\dagger)]\})\biggr) + \mathrm{H.c.}\\
\end{split}
\end{equation}

 Each of the terms in this sum (over $k$) commutes with the other terms, so we can consider the dynamics of each term separately, writing $H^s_{LI}=H^s_{LI,0}+\sum_k H^s_{LI,k}$. The motion-independent term $H^s_{LI,0}$ represents an oscillating Stark shift whose effect can be accounted for either through precise timing, in which case it leads to a phase shift (this would be difficult), or through an adiabatic ramp up and down of the laser pulses (in which case the accumulated phase can be made negligible). We now move to the interaction frame of the motion ($a_k\rightarrow a_k e^{-i \Omega_k t}$, $a_k^\dagger\rightarrow a_k^\dagger e^{i \Omega_k t}$) and define mode detunings $\delta_k=\Omega_k-2\omega_\Delta$.
We neglect counter-rotating terms to obtain
\begin{equation}
\begin{split}
    H^{s,m}_{LI,k}&= \sum_j \frac{1}{2} \biggl( [\Delta_{0,\Delta}(x_{0,j})\sigma_{z,j}+\Delta_{0,\Sigma}(x_{0,j}) ]\\
    &\times \exp(i\{-2\omega_\Delta t - \phi_{0,j}\}) \\
    &\times \{i \eta_{kj} (a_k e^{-i \Omega_k t}+a_k^\dagger e^{i \Omega_k t})\}\biggr) + \mathrm{H.c.}\\ \\
    &= a_k^\dagger \exp(i \delta_k t)  \sum_j \biggl( \frac{i}{2} \exp(-i \phi_{0,j}) \eta_{kj} \\
    &\times [\Delta_{0,\Delta}(x_{0,j})\sigma_{z,j}+\Delta_{0,\Sigma}(x_{0,j}) ]\biggr) + \mathrm{H.c.}\\
\end{split}
\end{equation}
To more easily understand the dynamics we write out all four components of this Hamiltonian for two ions (diagonal in the $\sigma_z$ basis). We define
\begin{equation}
\label{eqn:FKs}
\begin{split}
    F_{k,\Delta,+}&=\frac{i}{2} [\eta_{k1} \exp(-i\phi_{0,1}) \Delta_{0,\Delta}(x_{0,1})\\
    &\quad+ \eta_{k2} \exp(-i\phi_{0,2}) \Delta_{0,\Delta}(x_{0,2}) ]\\
    F_{k,\Delta,-}&=\frac{i}{2} [\eta_{k1} \exp(-i\phi_{0,1}) \Delta_{0,\Delta}(x_{0,1})\\
    &\quad - \eta_{k2} \exp(-i\phi_{0,2}) \Delta_{0,\Delta}(x_{0,2}) ]\\
    F_{k,\Sigma,+}&=\frac{i}{2} [\eta_{k1} \exp(-i\phi_{0,1}) \Delta_{0,\Sigma}(x_{0,1})\\
    &\quad + \eta_{k2} \exp(-i\phi_{0,2}) \Delta_{0,\Sigma}(x_{0,2}) ]\\
\end{split}
\end{equation}
in terms of which
\begin{equation}
\begin{split}
    H^{s,m}_{LI,k}(\uparrow\uparrow)&= (F_{k,\Delta,+} + F_{k,\Sigma,+}) \exp(i \delta_k t) a_k^\dagger + \mathrm{H.c.}\\
    H^{s,m}_{LI,k}(\uparrow\downarrow)&= (F_{k,\Delta,-} + F_{k,\Sigma,+}) \exp(i \delta_k t) a_k^\dagger + \mathrm{H.c.}\\
    H^{s,m}_{LI,k}(\downarrow\uparrow)&= (-F_{k,\Delta,-} + F_{k,\Sigma,+}) \exp(i \delta_k t) a_k^\dagger + \mathrm{H.c.}\\
    H^{s,m}_{LI,k}(\downarrow\downarrow)&= (-F_{k,\Delta,+} + F_{k,\Sigma,+}) \exp(i \delta_k t) a_k^\dagger + \mathrm{H.c.}\\
\end{split}
\end{equation}
This is a spin-dependent force Hamiltonian, and it drives the usual closed trajectories in motional phase space for appropriately chosen pulse durations and detunings. Assuming a fixed detuning and intensity, after an interval $\tau_g=n \frac{2\pi}{\delta_k}$ the mode has undergone $n$ circles in phase space and the state has acquired a geometric phase
\begin{equation}
\begin{split}
    \Phi^\prime(\uparrow\uparrow)&=\sign(\delta_k) \frac{2\pi n}{\delta_k^2} |(F_{k,\Delta,+} + F_{k,\Sigma,+})|^2\\
    \Phi^\prime(\uparrow\downarrow)&=\sign(\delta_k)\frac{2\pi n}{\delta_k^2} |(F_{k,\Delta,-} + F_{k,\Sigma,+})|^2\\
    \Phi^\prime(\downarrow\uparrow)&=\sign(\delta_k)\frac{2\pi n}{\delta_k^2} |(-F_{k,\Delta,-} + F_{k,\Sigma,+})|^2\\
    \Phi^\prime(\downarrow\downarrow)&=\sign(\delta_k)\frac{2\pi n}{\delta_k^2} |(-F_{k,\Delta,+} + F_{k,\Sigma,+})|^2.\\
\end{split}    
\end{equation}

We note that the interaction as derived cannot be used in the naive way to achieve a phase gate with a single pulse, because all four states in general acquire different phases. Rather we imagine a scheme with two pulses separated by a spin echo, so that the total phase acquired by $\ket{\uparrow\uparrow}$ and by $\ket{\downarrow\downarrow}$ are equal, and the phase acquired by $\ket{\uparrow\downarrow}$ and by $\ket{\downarrow\uparrow}$ are also equal. This has the added benefit of cancelling the quasi-static Stark shifts neglected earlier as well as increasing the resilience of the gate to slow qubit frequency drifts. In this case the acquired geometric phase is
\begin{equation}
\begin{split}
    \Phi(\uparrow\uparrow)&=\sign(\delta_k)\frac{2\pi}{\delta_k^2} (|(F_{k,\Delta,+} + F_{k,\Sigma,+})|^2\\
    &\quad+ |(-F_{k,\Delta,+} + F_{k,\Sigma,+})|^2)\\
    \Phi(\uparrow\downarrow)&=\sign(\delta_k)\frac{2\pi}{\delta_k^2} (|(F_{k,\Delta,-} + F_{k,\Sigma,+})|^2\\
    &\quad+ |(-F_{k,\Delta,-} + F_{k,\Sigma,+})|^2)\\
    \Phi(\downarrow\uparrow)&=\sign(\delta_k)\frac{2\pi}{\delta_k^2} (|(F_{k,\Delta,-} + F_{k,\Sigma,+})|^2\\
    &\quad+ |(-F_{k,\Delta,-} + F_{k,\Sigma,+})|^2)\\
    \Phi(\downarrow\downarrow)&=\sign(\delta_k)\frac{2\pi}{\delta_k^2} (|(F_{k,\Delta,+} + F_{k,\Sigma,+})|^2\\
    &\quad+ |(-F_{k,\Delta,+} + F_{k,\Sigma,+})|^2).\\
\end{split}    
\end{equation}
To implement the ideal phase gate we want
$\left|\Phi(\uparrow\uparrow)-\Phi(\uparrow\downarrow)\right|=\frac{\pi}{2}$, so
\begin{equation}
    \left|\frac{4\pi \eta_{k1} \eta_{k2} \cos(\phi_{0,2}-\phi_{0,1}) \Delta_{0,\Delta}(x_{0,1}) \Delta_{0,\Delta}(x_{0,2})}{\delta_k^2} \right| = \frac{\pi}{2}.
\end{equation}
For a gate at fixed detuning incorporating two loops in phase space within a duration $\tau_g$, the appropriate detuning is $\delta_k=\frac{4\pi}{\tau_g}$, which leads to
\begin{equation}
    \tau_g=\sqrt{\frac{2\pi^2}{\left|\eta_{k1} \eta_{k2} \cos(\phi_{0,2}-\phi_{0,1}) \Delta_{0,\Delta}(x_{0,1}) \Delta_{0,\Delta}(x_{0,2})\right|}}.
\end{equation}
This duration is minimized when the two ions are spaced by a half-integer multiple of the lattice wavelength: $\phi_{0,2}-\phi_{0,1}=m \pi$.

Note that the chosen detuning $\delta_k$ and gate duration $\tau_g$ do not in general close the trajectories in phase space for the spectator mode. For example, if the gate is performed with a detuning close to the two-ion breathing mode, the center-of-mass-mode phase-space trajectories may not be closed. However, given the much larger detuning from the this mode, one can choose from several possible gate durations and detunings close to the desired values for which both sets of trajectories are closed.

Gate sensitivity to mode-frequency variations is often reduced through appropriate Walsh modulation of the OTDF beam phase difference \cite{hayes_coherent_2012}. In the simplest situation this is achieved automatically when $F_{k,\Sigma,+}=0$. This will be the case only in special scenarios, such as for two ions of the same species, illuminated at the same intensity, spaced by an integer lattice wavelength multiple, driven near the breathing mode ($\eta_{k1}=-\eta_{k2}$) (see Eq.~\ref{eqn:FKs}). Nevertheless, Walsh modulation can still achieve the desired effect in more general situations via a sequence consisting of four pulses: reverse the OTDF phase between pulses one and two, flip the spins between pulses two and three, and again reverse the phase between pulses three and four.

\bibliography{phase_gate}
\end{document}